\documentclass[conference]{IEEEtran}
\usepackage{cite}

\ifCLASSINFOpdf
  \usepackage[pdftex]{graphicx}
\else
\fi
\usepackage{amsmath}
\usepackage{xcolor}

\begin{document}
%
\title{An Accelerated Testing Approach for Automated Vehicles with Background Traffic Described by Joint Distributions}

\author{\IEEEauthorblockN{Zhiyuan Huang}
\IEEEauthorblockA{Department of Industrial and\\Operations Engineering\\
University of Michigan\\}
\and
\IEEEauthorblockN{Henry Lam}
\IEEEauthorblockA{Department of Industrial Engineering\\
and Operations Research\\
Columbia University\\}
\and
\IEEEauthorblockN{Ding Zhao}
\IEEEauthorblockA{Department of Mechanical Engineering\\
University of Michigan\\
Corresponding author: zhaoding@umich.edu}
}


%


\maketitle

\begin{abstract}
This paper proposes a new framework based on joint statistical models for evaluating risks of automated vehicles in a naturalistic driving environment. The previous studies on the Accelerated Evaluation for automated vehicles are extended from multi-independent-variate models to joint statistics. The proposed toolkit includes exploration of the rare event (e.g. crash) sets and construction of accelerated distributions for Gaussian Mixture models using Importance Sampling techniques. Furthermore, the monotonic property is used to avoid the curse of dimensionality introduced by the joint distributions. Simulation results show that the procedure is effective and has a great potential to reduce the test cost for automated vehicles.
\end{abstract}


%
\IEEEpeerreviewmaketitle

\section{Introduction}

As Automated Vehicles (AVs) entering the market, it is important to form a systematic evaluation procedure for safety testing. The Naturalistic Field Operational Test (N-FOT) \cite{FESTA-Consortium2008}, which is adopted by many companies, is inefficient due to the rareness of safety critical scenarios.

In \cite{Zhao2016AcceleratedTechniques}, we proposed a brand new concept, the Accelerated Evaluation, and used the procedure to evaluate Automated Vehicles crash risks in a natural driving environment. We mainly explored the application on models of interaction between AVs and human-controlled vehicles. In \cite{Zhao2015j,Zhao2016g}, we modeled the frontal crash with a lead vehicle using single variable Gaussian process. We also explored different approaches to tackling the evaluation of lane change scenario \cite{Zhao2016AcceleratedTechniques,huang2017accelerated}. 

In previous work, our studies on the lane change scenario are based on the independence of random variables. We managed to decompose the density distribution and model the randomness using single variate distributions. In \cite{Zhao2016AcceleratedTechniques}, we used single parametric distribution to model the variables and we proposed the piecewise mixture distribution \cite{Huang2016UsingScenario} to increase the accuracy of model and evaluation efficiency. Cross Entropy method was used to optimize the importance sampling distribution. While the independence of variables is concluded from data observation, it is not strictly proved. Ignoring the dependence between variables might lead to some estimation error. Moreover, the independence of variables is not a general condition. The requirement of independence restraints the application domain of the Accelerated Evaluation procedure.

In this paper, we proposed an Accelerated Evaluation procedure using Gaussian Mixture Model (GMM). Gaussian Mixture Model is suitable and flexible for multivariate data. Data fitting using the Gaussian Mixture Model not only enabled us to capture the dependency between variables, but also provide accuracy if the number of mixture components is appropriate. 

%
%
%

Due to the numbers of parameters, using the Cross Entropy method to construct Importance Sampling distribution can be inefficient. The Cross Entropy method requires solving a stochastic optimization problem for several iterations. When the number of parameters is large, we need more samples to obtain a reasonable solution. This largely increases the required number of samples in the overall procedure. To avoid the inefficiency of the Cross Entropy method, We develop a new procedure based on the monotonicity property of rare event set, which we define in this paper. We derive a rare event set learning procedure to construct Importance Sampling distribution using both monotonicity property of rare-event sets and efficiency of change of measure for Gaussian distributions. 


We note that classification techniques such as support vector machine does not suit this problem, because the construction of Importance Sampling distribution requires specific properties for the form of the approximated rare-event set. While the proposed learning procedure obtains approximation of rare-event set with the required properties.

Section \ref{sec:lane} reviews the setting of the lane change scenario. In Section \ref{sec:gmm}, we review the truncated Gaussian Mixture Model fitting. The definition of monotonicity rare event sets and the set learning algorithm are present in Section \ref{sec:monotone}. We introduce how to construct an efficient Important Sampling distribution for Gaussian Mixture Model in Section \ref{sec:is} and show the specific procedure for monotonic rare events in \ref{sec:procedure}. We present the simulation results for the lane change scenario in Section \ref{sec:result}. Section \ref{sec:conclusion} concludes the paper.

\section{The Lane Change Scenario}\label{sec:lane}

The scenario studied in this paper is the defined as the following: a human-controlled vehicle driving in front of an automated vehicle start to cut into the lane where the automated vehicle is running. We want to evaluate the risk of a crash in this scenario. Since the scenario is initialized by humans, we can take it as a random environment. Then we can use the Automated Vehicle model to simulate the output (crash or not) for a given initial condition.

To model the initial condition of lane changes, we extracted data from the Safety Pilot Model Deployment (SPMD) database \cite{Bezzina2014}. The database includes over 2 million miles of vehicle driving data collected from 98 cars over 3 years. we identify 403,581 lane change events and use 173,692 events with a negative range rate to build statistical models. Three key variables can capture the effects of gap acceptance of the lane changing vehicle: velocity of the lead vehicle $v$, range to the lead vehicle $R$ and time to collision $TTC$. $TTC$ was defined as:
%
%
%
%
\begin{equation}
	TTC=- \frac{R}{\dot{R}},
\end{equation}
where $\dot{R}$ is the relative speed.  
\begin{figure}[t]
	\centering
	\includegraphics[width=0.8\linewidth]{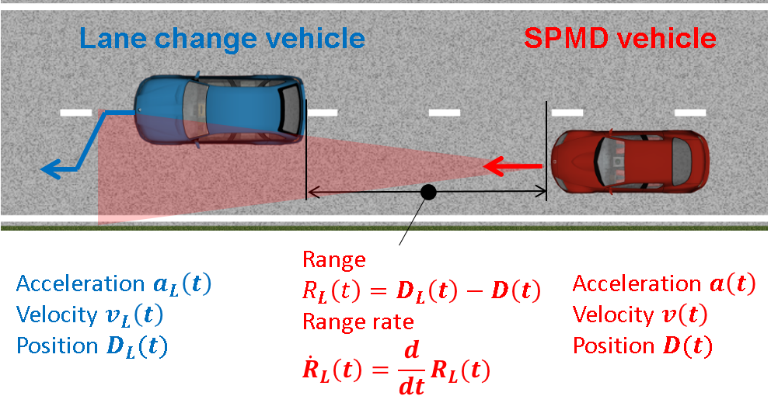}
	\caption{Lane change data collected by SPMD vehicle.}
	\label{fig:lane}
\end{figure}

%
%
%
%
The simulation of Automated Vehicle model is based on Adaptive Cruise Control (ACC) and Autonomous Emergency Braking (AEB) \cite{Ulsoy2012a} regarding the surrounding environment. With an initial condition, the simulator returns an output of the scenario. We consider as an event indicator function $I_\varepsilon(x)$ that returns $1$ (crash) or $0$ (safe) depending on the event of interest.

\section{Gaussian Mixture Model Fitting}\label{sec:gmm}

In this section, we review the truncated Gaussian Mixture Model (GMM) fitting studied by Lee and Scott \cite{Lee2012b}. 

The fitting of Gaussian Mixture Model is well studied and generally used. In the lane change scenario, the data need to be considered as truncated. For instance, the range of two cars will not be negative. For this reason, we use the truncated GMM to model the initial condition.

For a $d$ dimensional dataset with truncated as a hyper-rectangle with vertices $s$ and $t$, observations $y^n$ satisfies $s\leq y^n \leq t$. Note that we can have $s_i=-\infty$ or $t_i=\infty$, which indicates the $i$th coordinate is not truncated below or above.

For a truncated GMM with $K$ components, we use observation to estimate parameters in \begin{equation}
g(y)=\sum_{k=1}^{K} \eta_k g_k(y),
\end{equation}
where $\eta_k$ is the mixing weights, $\sum_{k=1}^{K} \eta_k=1$, $g_k$ is truncated Gaussian with support $[s,t]$, mean $\mu_k$ and covariance $\Sigma_k$. Similarly as the vanilla GMM fitting, we use an Expectation-Maximization (EM) algorithm to estimate the parameters. With a proper initial value for the estimated parameters, we use the following algorithm to iterate for converged estimators.

The E-step is:
\begin{equation}
\langle z_k^n \rangle =\frac{\eta_k g_k(y^n)}{\sum_{l} \eta_l g_l(y^n)},
\end{equation}
where we define $\langle z_k^n \rangle = P(z_k^n=1|y^n)$ to denote the probability that $y^n$ is generated from the $k$th component.

The M-step is also similar to the ordinary GMM fitting except some correction terms:
\begin{equation}
\hat{\eta}_k =\frac{1}{N} \sum_n \langle z_k^n \rangle ,
\end{equation}
\begin{equation}
\hat{\mu}_k =\frac{\sum_n \langle z_k^n \rangle y^n}{\sum_n \langle z_k^n \rangle} - m_k ,
\end{equation}
\begin{equation}
\hat{\Sigma}_k =\frac{\sum_n \langle z_k^n \rangle (y^n-\hat{\mu}_k)(y^n-\hat{\mu}_k)^T}{\sum_n \langle z_k^n \rangle}+ H_k ,
\end{equation}
where\begin{equation}
m_k=\mathcal{M}^1(0,\Sigma_k;[s-\mu_k,t-\mu_k]),
\end{equation}
\begin{equation}
H_k=\Sigma_k-\mathcal{M}^2(0,\Sigma_k;[s-\mu_k,t-\mu_k]).
\end{equation}
$\mathcal{M}^1(\mu,\Sigma;[a,b])$ and $\mathcal{M}^2(\mu,\Sigma;[a,b])$ denote the first and second moment of truncated Gaussian distribution with truncating vertices $a,b$, mean $\mu$ and covariance $\Sigma$.

The choice of number of components $K$ can be determined by some criteria for goodness of fitting, for example, we can use the Bayesian Information Criterion (BIC).

\section{Importance Sampling Distribution for Gaussian Mixture Models}\label{sec:is}
In this section, we first review the concept of Importance Sampling and some known Importance Sampling schemes for rare events with Gaussian distribution. We derive the scheme for GMM based on these background knowledge.
\subsection{Importance Sampling for Gaussian Distribution}
Importance Sampling is a technique to reduce the variance in simulation. 

Consider a random vector $x$ with distribution $F$ and a rare event set $\varepsilon \subset \Omega$ on sample space $\Omega$. Our goal is to estimate the probability of the rare event \begin{equation}
{P}(X \in \varepsilon)=E[I_\varepsilon(X)]=\int I_\varepsilon(x) dF,
\end{equation} where the event indicator function is defined as\begin{equation}
	I_\varepsilon(x)=\begin{cases} 1 & x \in \varepsilon.\\
0 & otherwise.\end{cases}
\end{equation}

The crude Monte Carlo using the sample mean of $I_\varepsilon(x)$ \begin{equation}
	\hat{P}(X \in \varepsilon) = \frac{1}{N} \sum_{n=1}^N I_\varepsilon(X_n),
\end{equation}
where $X_i$'s are drawn from distribution $F$.

The Importance Sampling \cite{Bucklew2004a} technique is derived from\begin{equation}
	E[I_\varepsilon(X)]=\int I_\varepsilon(x) dF = \int I_\varepsilon(x) \frac{dF}{dF^*} dF^* ,
\end{equation}
which gives the estimator using the sample mean of the above expectation (use ref) \begin{equation}
	\hat{P}(X \in \varepsilon) = \frac{1}{N} \sum_{n=1}^N I_\varepsilon(X_n) \frac{dF}{dF^*},
\end{equation}
where $X_i$'s are generated from $F^*$, which has the same support with $F$. We note that this is an unbiased estimation of ${P}(X \in \varepsilon) $. By appropriately selecting $F^*$, the evaluation procedure obtains an estimation with smaller variance. $F^*$ is called the IS distribution. We define the likelihood function \begin{equation}
L(x)=\frac{dF(x)}{dF^*(x)}.
\end{equation}

The analysis of asymptotic efficiency \cite{Asmussen2007StochasticAnalysis,Bucklew2004a} is a benchmark to determine whether an IS distribution is proper. Let $Z=L(x) I_\varepsilon(x)$ be an IS estimator, it is  efficient if \begin{equation}
\lim_{\varepsilon \rightarrow \infty} \frac{\log \left( E_{F^*} [Z^2] \right)}{\log \left( E_{F^*}[Z] \right)} \leq 2,
\end{equation}
where $\varepsilon \rightarrow \infty$ denotes that the rare event set $\varepsilon$ diverges to $\infty$ in a suitable sense (e.g., $\inf_{x\in \varepsilon}  \| x\|_2 \rightarrow \infty $).

When $x$ follows Gaussian distribution with mean $\mu$ and covariance matrix $\Sigma$ and the rare event set $\varepsilon$ satisfies the convexity assumption, there is a simple scheme that obtains an efficient IS distribution. Here, we introduce the scheme with assumptions satisfied in the lane change scenario.

For a convex set rare event set $\varepsilon$, we define the dominating point of $\varepsilon$ on $\phi(\cdot;\mu,\Sigma)$ to be \begin{equation}
	a^*=\arg \max_{a\in \varepsilon} \phi(a;\mu,\Sigma),
\end{equation}
where $\phi(\cdot;\mu,\Sigma)$ is the density function for Gaussian distribution with mean $\mu$ and covariance matrix $\Sigma$. The dominating point contributes the highest density among all points in $\varepsilon$. By shifting the mean $\mu$ of the Gaussian distribution to $a^*$, we can obtain an IS distribution that provides a effecient estimator for $P(x \in \varepsilon)$ \cite{sadowsky1996monte,dieker2005asymptotically}.

\subsection{IS Scheme for Gaussian Mixture Model and Union of Convex Rare Event Sets}
Based on the IS scheme for Guassian distribution on convex rare event set, we propose an IS scheme for Gaussian Mixture Model on the union of convex rare event sets. 
  
Assume $x$ follows a Gaussian Mixture Model with $k$ components, the density of $x$ is \begin{equation}
	f(x)=\sum_{i=1}^{k}p_i \phi(x;\mu_i,\Sigma_i).
	\label{eq:gmm_density}
\end{equation}
The rare event set $\varepsilon$ is consisted with $l$ convex sets, we denote as $\varepsilon = \cup_{j=1}^{l} \varepsilon_j$.

For each convex set $\varepsilon_j$, we find the dominating point of $\varepsilon_j$ on the Gaussian component $\phi(x;\mu_i,\Sigma_i)$ by \begin{equation}
	a_{ij}^*=\arg \max_{a\in \varepsilon_j} \phi(a;\mu_i,\Sigma_i).
\end{equation}
We propose to use the IS distribution as following:\begin{equation}
	f^*(x)=\sum_{i=1}^{k} \sum_{j=1}^{l} p_i q_j \phi(x;a_{ij},\Sigma_i),
\end{equation}
where the $q_j$ can be arbitrary positive number that satisfies $\sum_{j=1}^{l} q_j=1$. We use $q_j={1}/{l}$. 

\begin{figure}[t]
	\centering
	\includegraphics[width=\linewidth]{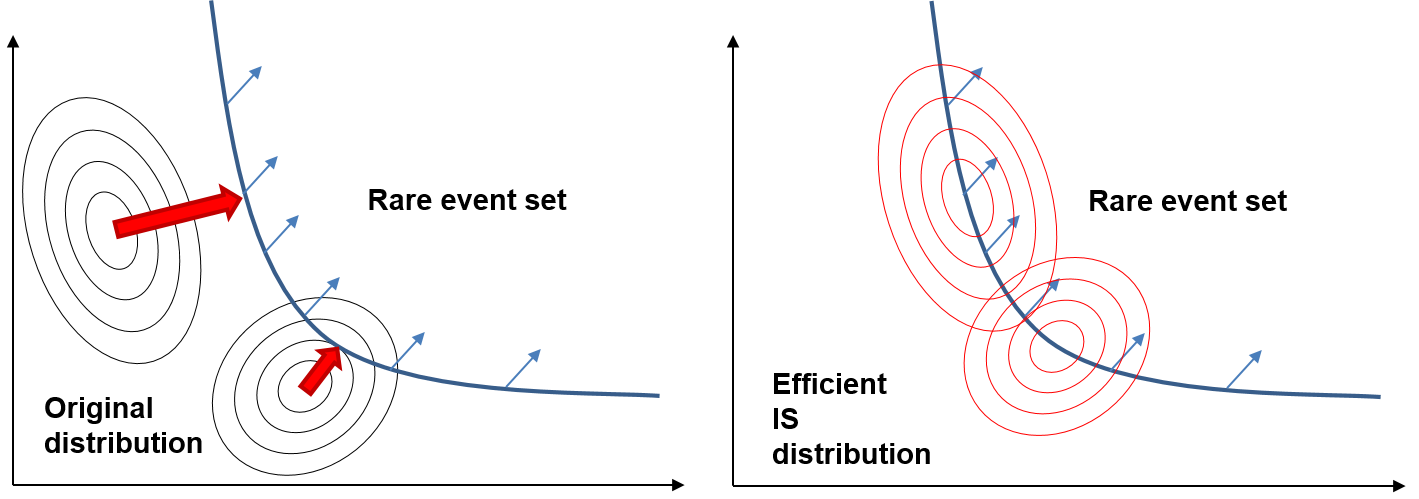}
	\caption{Scheme of constructing accelerated distribution for Gaussian Mixture Model with convex rare event set.}
	\label{fig:stgy1}
\end{figure}

\begin{figure}[t]
	\centering
	\includegraphics[width=\linewidth]{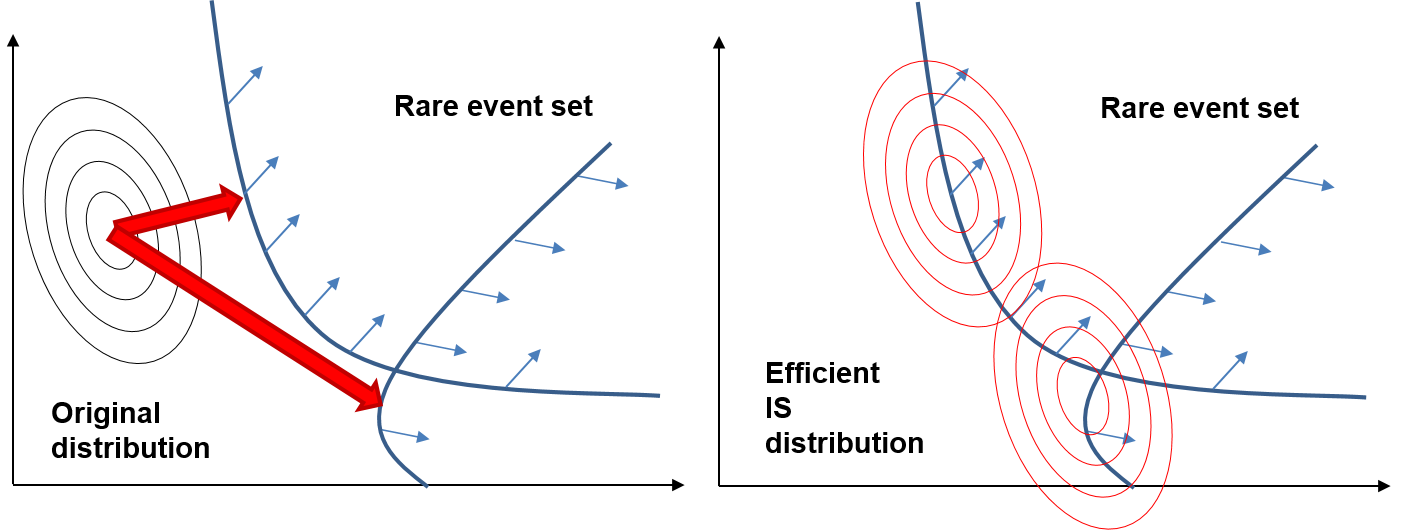}
	\caption{Scheme of constructing accelerated distribution for single Gaussian Model with union of convex rare event sets.}
	\label{fig:stgy2}
\end{figure}

Note that this scheme can be viewed as the combination of two basic schemes shown is Fig. \ref{fig:stgy1} and Fig. \ref{fig:stgy2}. For each mixture component, we find the dominating points for the convex sets in the union of sets and form a IS distribution based on the dominating points. We use the mixture of IS distributions of all mixture components as the IS distribution for the model.

%
%

\section{Monotonic Rare Event Sets Learning}\label{sec:monotone}
In this section, we first define the monotonicity for rare event sets. Then we propose a learning algorithm to obtain an outer approximation set, which contains the rare event set, and an inner approximation set, which is a subset of the rare event set, based on the monotonicity property. The approximation sets we obtain are unions of convex sets, which suit in the IS scheme we proposed in Section \ref{sec:is}.

\subsection{Definition of Monotonic Rare Event Sets}
For a rare event set $\varepsilon$ on $d$ dimensional space, if $x_1 \in \varepsilon$ and $x_1 \leq x_2$ implies that $x_2 \in \varepsilon$, we define the set $\varepsilon$ non-decreasing. We note that non-increasing set can be defined by flipping inequality as $x_1 \geq x_2$. Both non-decreasing and non-increasing set are defined to be monotonic.

For example, in the lane change scenario, if a crash occurs for initial condition $(v, TTC, R)$, then if any of these variable is smaller, we can determine that a crash will happen. The set for crash is non-increasing. This can be explained intuitively, because a smaller $v$, $TTC$ and $R$ means that there is less room for the following Automated Vehicle to make adjustments.

\begin{figure}[t]
	\centering
	\includegraphics[width=0.8\linewidth]{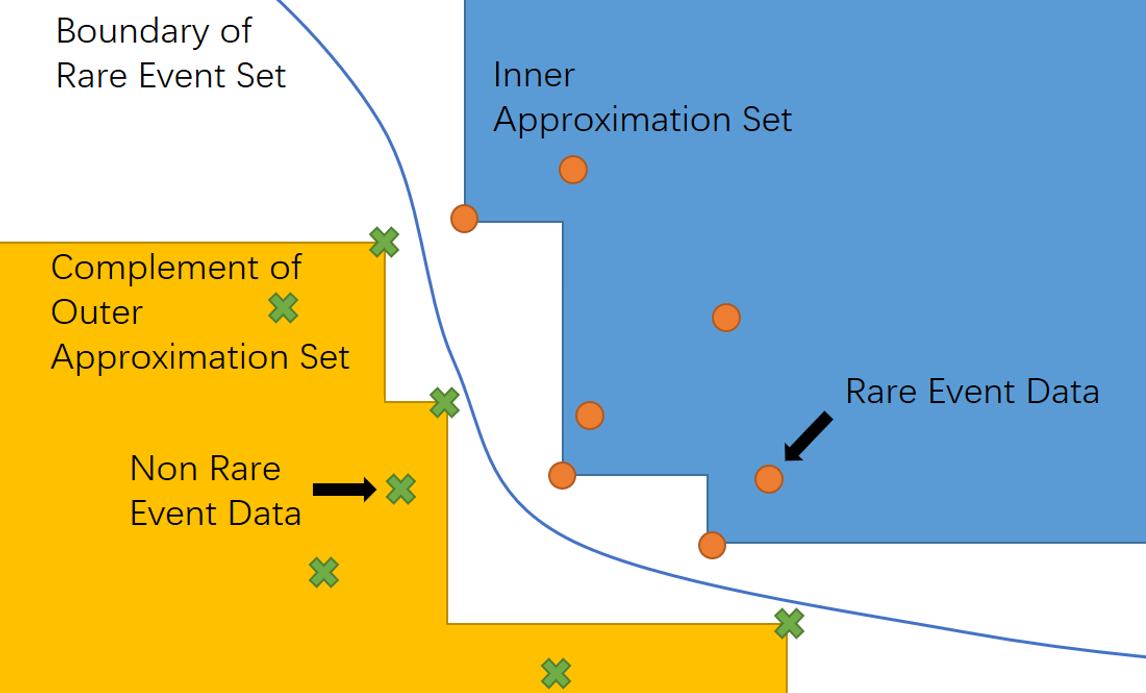}
	\caption{An illustration of the monotonic set learning.}
	\label{fig:mono}
\end{figure}
%
%

\subsection{Learning Algorithm}

Since our goal for learning the approximation set of rare event is to construct an IS distribution, we want the approximated set to satisfy the assumption in the scheme we proposed in Section \ref{sec:is}. Therefore, we want the approximation set to be an union of convex sets.

Let us consider a non-decreasing rare event set $\varepsilon $ on a $d$ dimensional space. We denote $\mathcal{S}_1=\{a^1,...,a^{n_1}\}$ as the observed data in the rare event set and $\mathcal{S}_0=\{b^1,...,b^{n_0}\}$ as the observed data not contained by the rare event set. We use $\mathcal{S}_1$ to construct an inner approximation set $\underline{\varepsilon}$ and use $\mathcal{S}_0$ to construct an outer approximation set $\bar{\varepsilon}$. 

For each data point $a_i$ in $\mathcal{S}_1$, we construct a set $\mathcal{I}_i=\{x:x \geq a_i\}$. By the definition of non-decreasing set, we have $\mathcal{I}_i \subset \epsilon$. Therefore, we have $\underline{\varepsilon}= \cup_{i=1}^{n_1} \mathcal{I}_i \subset \epsilon$ as the inner approximation set of $\varepsilon $.

 For each data point $b^j$ in $\mathcal{S}_0$, we construct sets $\mathcal{O}_{jk}=\{x:x_k \geq b_k^j\} $ for $k=1,...,d$, where $x_k$ denotes the $k$th element in $x$ and $b_k^j$ denotes the $k$th element in $b^j$. We have $\epsilon \subset \cup_{k=1}^{d} \mathcal{O}_{jk}$, which indicates that $\epsilon \subset \cap_{j=1}^{n_0} \cup_{k=1}^{d} \mathcal{O}_{jk}$. We define $\bar{\varepsilon} = \cap_{j=1}^{n_0} \cup_{k=1}^{d} \mathcal{O}_{jk}$ as the outer approximation set of $\varepsilon $. We can express $\bar{\varepsilon} $ as union of convex sets $\cup_{m_1=1}^d...\cup_{m_{n_0}=1}^d \left( \cap_{k=1}^{n_0} \mathcal{O}_{km_k}\right)$. Fig. \ref{fig:mono} illustrates the key idea of this learning procedure.

We note that $\mathcal{I}_i$'s and $\mathcal{O}_{jk}$'s are convex, the approximation sets are unions of convex sets. We also note that using the minimums of $\mathcal{S}_1$ or maximums of $\mathcal{S}_0$ (also known as Pareto fronts) of the dataset provides the same inner or outer approximation set. Here, minimums are defined as points that are greater in all dimensions than no other points; whereas maximums are defined as points that are smaller in all dimensions than no other points.

Additionally, to use the result for non-decreasing sets, we can simply flip all coordinates of data (take negative value) when we have a non-increasing set. For sets that is non-decreasing on some of the coordinates and non-increasing on the rest coordinates, we can flip the value of data those non-increasing coordinates. The approximation sets obtained will still be unions of convex sets after flipping any coordinate.

We note that we can directly obtain a lower probability bound and an upper probability bound for the rare event probability.

\section{IS Distribution Construction Scheme for A Gaussian Mixture Model of A Monotonic Rare Event Set} \label{sec:procedure}

Combining the discussion in Sections \ref{sec:is} and \ref{sec:monotone}, we propose an iterative procedure that provides IS distributions for a Gaussian Mixture Model with a monotonic rare event set.

Here, we consider a non-decreasing rare event set $\varepsilon$ on $d$ dimensional space and the variable vector $x$ is generated from a $k$ components GMM with density (\ref{eq:gmm_density}). To clarify the notations, for each Gaussian component $i$, we construct a set of dominating points $\mathcal{A}_I^i$ using the inner approximation set $\underline{\varepsilon}$ and $\mathcal{A}_O^i$ using the outer approximation set $\bar{\varepsilon}$ for $i=1,...,k$. $|\mathcal{A}|$ denotes the number of elements in the set $\mathcal{A}$. $\underline{\varepsilon}$ and $\bar{\varepsilon}$ are constructed based on $\mathcal{S}_1$, which contains the minimums of observed data in $\varepsilon$, and $\mathcal{S}_0$, which contains the maximums of observed data that not in $\varepsilon$. The procedure iterates to update these sets, which will provide two IS distributions $f^*_I$ and $f^*_O$ based on the inner approximation and the outer approximation of $\varepsilon$ respectively.

The procedure is presented as the following:
\begin{enumerate}
\item Initialize $A^i_I=A^i_O=\{\mu_i\}$ and $\mathcal{S}_1=\mathcal{S}_0=\emptyset$.
\item Construct the sampling distribution\begin{equation}
f^*_I(x)=\sum_{i=1}^{k} \sum_{a \in \mathcal{A}^i_I} p_i \frac{1}{|\mathcal{A}^i_I|} \phi(x;a,\Sigma_i)
\end{equation} and \begin{equation}
f^*_O(x)=\sum_{i=1}^{k} \sum_{a \in \mathcal{A}^i_O} p_i \frac{1}{|\mathcal{A}^i_O|} \phi(x;a,\Sigma_i)
\end{equation} \label{step:start}

\item Sample $N$ data points $D=\{x_1,...,x_N\}$ from the density \begin{equation}
f(x)=\rho f^*_I(x) +(1-\rho) f^*_O(x), \label{eq:is_con}
\end{equation}
where $\rho$ is on $[0,1]$. We suggest to use $\rho= 1/2 \ I_{\{\mathcal{S}_1 \neq \{\mu_i\}\}}.$

\item    Input $D$ to the simulator $I_\varepsilon(x)$ and use the outcome to update $S_1$ and $S_0$. We add the new data points to $S_1$ and $S_0$ regarding the outcome of $I_\varepsilon(x)$. Then we discard non-minimum data points in $S_1$ and non-maximum data points in $S_0$.

\item For each Gaussian component $i$, we use each data points $a \in S_1$ to solve \begin{equation}
\max_x \ \phi(x;\mu_i,\Sigma_i)\ \text{subject to } x\geq a. \label{eq:opt_nominate_1}
\end{equation}
We obtain $n_1=|S_1|$ solutions and the solution set is our new $\mathcal{A}^i_I$.

\item For each Gaussian component $i$, we use each data points $b^j \in S_0$ to solve \begin{equation}
\max_x \ \phi(x;\mu_i,\Sigma_i)\ \text{subject to } x_{m_k}\geq b_{m_k}^k,\ k=1,...,n_0 \label{eq:opt_nominate_0}
\end{equation}
for $m_1=1,...,d$,  ..., $m_{n_0}=1,...,d$. $ x_m\geq b_m$ denotes the $m$th element of $x$ is greater or equals to the $m$th element of $b$.
We obtain $d^{n_0}=d^{|S_0|}$ solutions and the solution set is our new $\mathcal{A}^i_O$. \label{step:end}

\item Iterate from \ref{step:start}) to \ref{step:end}).

\end{enumerate}

The stop criterion can either be a maximum iteration number or a maximum number of elements in either set if reached. A recommended IS distribution is in the form of (\ref{eq:is_con}). Since the optimal $\rho$ is case-by-case, we need to observe the elements in $\mathcal{A}_I^i$'s and $\mathcal{A}_O^i$'s to decide the value of $\rho$. If there elements in $\mathcal{A}_I^i$'s and $\mathcal{A}_O^i$'s are very different (in the sense of variable values) or there are few elements in , we recommend to use $\rho=0$. If $\mathcal{A}_I^i$'s and $\mathcal{A}_O^i$'s are similar, then we use $\rho=0.5$. 

\section{Simulation Analysis on Lane Change Scenario}\label{sec:result}

In this section, we present the GMM fitting of the lane change model and use the IS distribution construction scheme proposed on the model. We show simulation results to justify the validity of the scheme.

\subsection{Truncated Gaussian Mixture Model Fitting}

\begin{figure}[t]
	\centering
	\includegraphics[width=\linewidth]{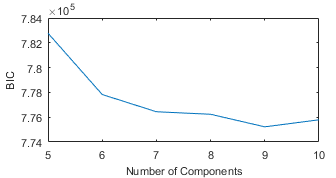}
	\caption{The BIC regarding to different number of components $K$.}
	\label{fig:bic}
\end{figure}

We use the EM algorithm in Section \ref{sec:gmm} to fit the lane change data $(v,TTC^{-1},R^{-1})$. The important part for GMM fitting is the selection of the number of components $K$. Fig. \ref{fig:bic} shows the BIC regards to different $K$. We could observe that we obtain a local minimum at $K=9$, where it means that the model with $K=9$ provides a balance between the number of parameters and the fitting. 

We note that the different scale of the variables might cause some numerical issues in implementing the algorithm presented in Section \ref{sec:gmm}, we can normalize (subtract by marginal mean and then divided by marginal standard deviation) the data before we fit the model.

\subsection{Importance Sampling Results}

The IS distribution construction scheme we proposed in this paper is based on the assumption of ordinary Gaussian Mixture Model. Since the distribution of variables in the lane change scenario is truncated, we need to make small modifications on the scheme.

Shifting the sampling mean to dominating points still works in for truncated Gaussian, because the truncated coordinate does not have asymptotic behavior. In this case, we only need to worry about those untruncated coordinates, which will be the same as untruncated Gaussian Distribution.

For the monotonic set learning part, note that if we directly use (\ref{eq:opt_nominate_1}) and (\ref{eq:opt_nominate_0}) for truncated variables, we might obtain in the infeasible region. We add the truncated boundary as constrains for these optimization problems.

After we obtain the dominating point sets $\mathcal{A}_I^i$'s and $\mathcal{A}_O^i$'s from the proposed algorithm, we observe that the elements in $\mathcal{A}_I^i$ and $\mathcal{A}_O^i$ are tend to be very different. Therefore, we use $\rho=0$ for the IS distribution.

\begin{figure}[t]
	\centering
	\includegraphics[width=\linewidth]{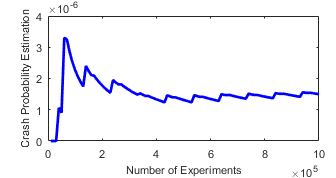}
	\caption{The crash probability estimation with increasing number of samples.}
	\label{fig:mean}
\end{figure}

\begin{figure}[t]
	\centering
	\includegraphics[width=\linewidth]{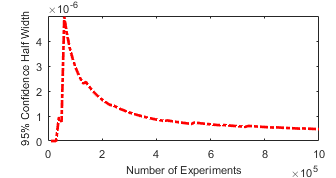}
	\caption{The confidence interval of the crash probability estimation with increasing number of samples.}
	\label{fig:ci}
\end{figure}

Fig. \ref{fig:mean} presents the estimated probability with different number of experiments. The probability converges around $1.15 \times 10^{-6}$. Fig. \ref{fig:ci} shows the $95\%$ confidence interval half width of the probability estimation. For $10^6$ samples, we have the $95\%$ confidence interval for the estimation as $(1.02\times 10^{-6},1.97\times 10^{-6})$. Using the formula for the standard deviation of the crude Monte Carlo estimation $\hat{P}(x \in \varepsilon)$ by \begin{equation}
std(\hat{P}(x \in \varepsilon))=\sqrt{\frac{\hat{P}(x \in \varepsilon)(1-\hat{P}(x \in \varepsilon))}{n}},
\end{equation} 
we can estimate that the crude Monte Carlo method requires about $2.56 \times 10^7$ samples to reach a confidence interval with a similar scale. The IS distribution constructed by the proposed algorithm increases the efficiency of the estimation by roughly $25$ times.

\section{Conclusion}\label{sec:conclusion}

This paper proposes using Gaussian Mixture Model to model stochastic variables in Automated Vehicle evaluating problems. We provide an algorithm for constructing Importance Sampling distribution based on the property of Gaussian Mixture model and monotonic rare events. The proposed algorithm can provide a valid Importance Sampling distribution.

A further direction for our research is to find a scheme to refine the elements in the dominating point sets.

\section*{Acknowledgment}

The authors acknowledge support from the University of Michigan Mobility Transformation Center: a project under grant number N021552.



%



\bibliographystyle{IEEEtran}
\bibliography{citations.bib}

\end{document}